\documentclass{Interspeech}

\interspeechcameraready 

\title{Segmental Attention Decoding With Long Form Acoustic Encodings}

\author[affiliation={1}]{Pawel}{Swietojanski}
\author[affiliation={1}]{Xinwei}{Li}
\author[affiliation={1}]{Mingbin}{Xu}
\author[affiliation={1}]{Takaaki}{Hori}
\author[affiliation={1}]{Dogan}{Can}
\author[affiliation={1}]{Xiaodan}{Zhuang}

\affiliation{}{Apple}{USA}
\email{\{pswietojanski, xinwei\_li2\}@apple.com}


\usepackage{comment}
\usepackage{amsmath}
\usepackage{amsfonts}
\usepackage{amssymb}  

\usepackage{graphicx}
\usepackage[acronym]{glossaries}
\usepackage[font=footnotesize]{subfig}
\usepackage{cite}
\usepackage{float}
\usepackage{multirow}
\usepackage{tikz}
\usepackage{amsthm}

\usepackage{multirow}
\usepackage{booktabs}

\usepackage{pifont}   
\usepackage{array}    
\usepackage{xspace}

\usepackage{placeins}
\usepackage{threeparttable}

\newcommand{\cmark}{\ding{51}}  
\newcommand{\xmark}{\ding{55}}  

\newtheorem{theorem}{Definition}

\newcommand{\sae}{SFE\xspace}
\newcommand{\lae}{LFE\xspace}

\newcommand{\mbase}{\textit{Ours.base}\xspace}
\newcommand{\msmall}{\textit{Ours.small}\xspace}

\newacronym{asr}{ASR}{Automatic Speech Recognition}
\newacronym{wer}{WER}{Word Error Rate}
\newacronym{etoe}{E2E}{End-to-end}
\newacronym{lm}{LM}{Language Model}
\newacronym{lstm}{LSTM}{Long Short-Term Memory}
\newacronym{bilstm}{BiLSTM}{Bidirectional Long Short-Term Memory}

\newcommand{\cf}{\textit{cf.~}}

\newcommand{\ie}{\textit{i.e.~}}

\newcommand\ignore[1]{}

\def \path{\bp C}

\newcommand{\bfp}{{\mathbf{p}}}

\newcommand{\calC}{{\mathcal{C}}}
\newcommand{\calD}{{\mathcal{D}}}
\newcommand{\calE}{{\mathcal{E}}}

\newcommand{\calH}{{\mathcal{H}}}

\newcommand{\calM}{{\mathcal{M}}}

\newcommand{\calX}{{\mathcal{X}}}
\newcommand{\calY}{{\mathcal{Y}}}

\begin{document}

\maketitle

\begin{abstract}
    
We address the fundamental incompatibility of attention-based encoder-decoder (AED) models with long-form acoustic encodings. AED models trained on segmented utterances learn to encode absolute frame positions by exploiting limited acoustic context beyond segment boundaries, but fail to generalize when decoding long-form segments where these cues vanish. The model loses ability to order acoustic encodings due to permutation invariance of keys and values in cross-attention. We propose four modifications: (1) injecting explicit absolute positional encodings into cross-attention for each decoded segment, (2) long-form training with extended acoustic context to eliminate implicit absolute position encoding, (3) segment concatenation to cover diverse segmentations needed during training, and (4) semantic segmentation to align AED-decoded segments with training segments. We show these modifications close the accuracy gap between continuous and segmented acoustic encodings, enabling auto-regressive use of the attention decoder.
\end{abstract}

\section{Introduction and Related Work}
\label{sec:intro}

Attention encoder decoder (AED) models~\cite{chorowski2015attention, chan2016listen} make a popular family of  end-to-end speech recognition models offering powerful auto-regressive capabilities, though long-form (LF) ASR with AED models remains a challenge~\cite{chiu2019comparison,Narayanan2020CascadedEF,fox2024updated}. Current approaches typically rely on external segmentation systems or different windowing techniques~\cite{radford2022whisper, hori20_interspeech, HoriMHR21} that artificially recreate boundary conditions. 
As such, attention decoder (AD) is often used in multi-pass settings instead where the first pass hypotheses are refined by AD~\cite{yao2021wenet, wang2021streaming}. 
Such rescoring, however, operates on n-best lists or discrete tokens, rather than full auto-regressive decoding limiting potential accuracy gains.
Most recent efforts related to LF ASR focus on transformer encoders with linearized attention~\cite{summarymixing, kim2023branchformer}, or alternative architectures~\cite{zhang2025mamba}.
\cite{carvalho25_interspeech} recently proposed long-form (LF) training, and evaluated different flavours of linear-time encoders on LF tasks. 
In this work we are interested in the AD inability to use LF encodings in auto-regressive manner. Issues with AD on LF data were identified before~\cite{Narayanan2020CascadedEF}, though not addressed to date to our knowledge.

Our contributions are fourfold: (1) We identify the permutation invariance problem in attention-based ASR for LF audio, demonstrating how short-form boundary effects hide this limitation during training. (2) To tackle (1), we propose modifications including cross-attention positional encodings and LF training. (3) We demonstrate that our approach achieves parity between segmented and long-form performance while maintaining competitive results against similarly-sized models. (4) Extend the model with first pass semantic segmentation capabilities, showing it outperforms voice-activity-detection (VAD).

\begin{figure}[htbp]
    \centering
    \includegraphics[width=0.65\textwidth, trim={18cm 10.5cm 3cm 4.5cm}, clip]{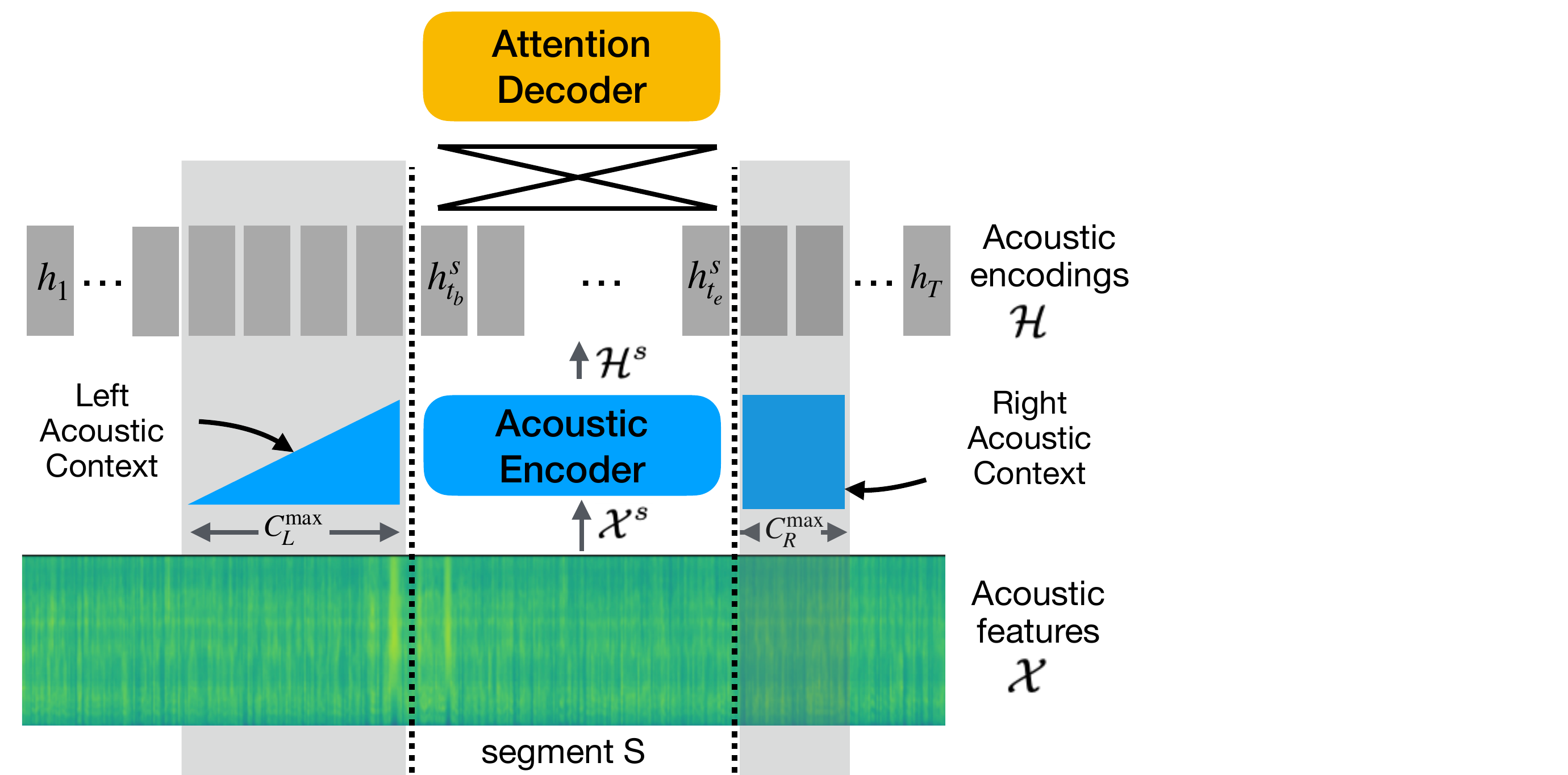}
    \caption{Illustration of \lae, left/right acoustic context and boundary conditions. Shaded areas denote regions crucial to \lae, absent during training and inference on segmented data.
    }
    \label{fig:lf}
    \vspace{-0.4cm}
\end{figure}

\section{Definitions}
\label{sec:method}

\subsection{Attention-based Encoder-Decoder Model}

The AED model converts speech audio to text using three main parts: an acoustic encoder, a cross-attention mechanism, and an autoregressive decoder~\cite{chan2016listen,chorowski2015attention}.

\textbf{Acoustic Encoder:} The encoder $\calH = \calE_{\theta}(\calX)$ takes a sequence of acoustic features $\calX= (x_1, ..., x_{T'})$ and converts them into sequence of acoustic encodings $\calH = (h_1, ..., h_T)$, with $T \leq T'$. Modern encoders use conformer~\cite{gulati2020conformer, rekesh2023fast} or transformer~\cite{vaswani2017attention, zhang2020transformer} blocks with convolutional downsampling.

\textbf{Attention Decoder:} The decoder $y_u = \calD_{\theta}(y_{u-1}, \calC_{<u}, \calH)$ produces output tokens $\calY = (y_1, ..., y_U)$ auto-regressively one token at a time. At each step $u$, the decoder combines three pieces of information: the previous token embedding $y_{u-1}$, its previous internal state $\calC_{<u}$ up to $u$, and acoustic encodings $\calH$. 

\textbf{Cross-Attention Mechanism:} $\calD_{\theta}$ interacts with encoder $\calE_{\theta}$ via cross-attention~\cite{vaswani2017attention} layers.
In AED, dot-product cross-attention queries come from decoder states $\calC$, while the keys and values are derived from the acoustic encodings $\calH$. 

\subsection{Long-Form Acoustic Encodings} \label{ssec:laf}

In this work, we investigate the relationship between the receptive field of the transformer-based encoder $\calE_{\theta}$ and the ability of the attention decoder $\calD_{\theta}$ to transcribe any segment $\calH^s \in \calH$, for encodings $\calH$ computed from an arbitrarily audio stream $\calX$.
Contrary to recurrent neural networks, the transformer architecture enables precise control of the neighbourhood information at each step. For example, acoustic encoding at time $t$, $h_t = \calE_{\theta}(\calX)_t$, may be derived from an arbitrary subset of features in $\calX$, as defined by the structure of the mask $\calM$ used in the self-attention layers of $\calE_{\theta}$. 
Definitions~\ref{def:lfa} and~\ref{def:lfs} specify the notions of long-form acoustic encoding, and long-form segment as relevant for this work:

\begin{theorem}[Long-Form Acoustic Encoding (\lae)] An acoustic encoding $h_t$ is an \lae when it satisfies:
\begin{equation}
h_t = \mathcal{E}_\theta(\mathcal{X})_t \quad \text{with} \quad C_L(t) = C_L^{\max}, \quad C_R(t) = C_R^{\max} \nonumber
\end{equation}
\noindent where $C_L^{\max}$ and $C_R^{\max}$ denote the maximum left/right context supported by the encoder $\mathcal{E}_\theta$ when processing inputs $\calX$.
\label{def:lfa}
\end{theorem}

\begin{theorem}[Long-form segment] A segment $s$ is considered long-form if each encoding $h^s_t \in \calH^s$ is an \lae encoding.
\label{def:lfs}
\end{theorem}

\noindent In other words, the encoding $h_t$ becomes \lae when computed with the full bidirectional context defined by the masking strategy of $\mathcal{E}_\theta$. Fig.~\ref{fig:lf} illustrates the crucial difference between short-form encodings (\sae) and \lae for any segment $S$, and its corresponding acoustic encodings $\calH^s = (h_{t_b}^s, ...., h_{t_e}^s)$. In particular, for segmented utterances (without shaded regions) $C_L(t_b)$ of $h_{t_b}^s$ and $C_R(t_e)$ of $h_{t_e}^s$ is $0$. In contrast, for \lae (with shaded regions) $C_L(t_b)=C_L^{\max}$ and $C_R(t_e)=C_R^{\max}$. Global attention makes a special case, as any encoding at any time can be anchored relative to the recording's edge encodings ($h_1$ and $h_T$), thus mitigating the LF AD issue. Global attention, however, is not always practical as i) the encoder may be expected to be streaming, and more importantly, ii) global attention for speech is known to be sensitive to duration mismatch between training and inference conditions~\cite{Likhomanenko2021_CAPE, Pan2022_SRU}, an issue that is likely to only be exacerbated in long-form scenarios with ever-growing left-context. In general, it is not straightforward to eliminate the short/long form mismatch for arbitrary length sequences, as the length of training utterances is usually more constrained than the length of unsegmented audio streams during inference.

In this work, we keep the left and right contexts of $\mathcal{E}_\theta$ fixed to a number of look-back and look-ahead conformer frames, similar to~\cite{zhang2020transformer}, ensuring robust estimation within this regime. 
Under this formulation, the expected context to obtain \lae depends on the finite number of $N$ "look-back" frames each self-attention layer sees in the preceding layer, the number of transformer layers $L$ and decimation factor $R$: $C^{\max}_L = L\cdot N\cdot R$ acoustic frames. For the right context, we use causal chunks~\cite{Chen2021} as a good trade-off between latency, accuracy and hardware utilization, thus $C^{\max}_R = M\cdot R$, where $M$ is the maximum number of frames within the chunk that self-attention layers can look at in the future. Additionally, similar to~\cite{var_masking} we train models with several chunk sizes so the latency/accuracy trade-off can be configured during inference. This context limited architecture works well in long-form applications.


\section{Proposed Modifications} \label{sec:data}

\subsection{Explicit Positional Encodings in Cross Attention} \label{ssec:pe}

The fundamental property of dot-product cross-attention is its permutation invariance~\cite{vaswani2017attention}. Given any ordering of inputs $\calH$, attention produces the same output, meaning it cannot distinguish between different acoustic orderings based solely on content, unless content itself encodes sequential order.

This property becomes problematic for continuous long form acoustic streams as AED is typically trained on segmented utterances where attention can implicitly learn to exploit edge effects as positional anchors where acoustic encodings at segment boundaries lack full bidirectional context (\cf Fig.~\ref{fig:lf}, without shaded areas). This creates distinctive patterns that help the decoder implicitly track the ordering of $\calH$. However, when the encoder processes enough acoustic signal as is the case in the long-form scenario (\cf Fig~\ref{fig:lf} with shaded areas), these edge cues vanish and all embeddings appear contextually similar, causing the issues with positional tracking. This limitation manifests as repeated transcription outputs and an inability of AD to emit end-of-sentence (EOS) tokens. 

To address permutation invariance, we add an absolute positional codes $\bfp$ to each segment $\calH^s$, \ie $\calH^s_p = \calH^s+\bfp$. This is on top of the usual positional information injected at the encoder inputs~\cite{vaswani2017attention, su2024rope}, as we also explicitly incorporate those at the \lae level $\calH$, before they are fed into cross-attention keys and values. This ensures attention weights reflect temporal position, while positional information specific to segment $s$ propagates to the decoder. Crucially, position indices are reset for each decoded segment rather than maintaining global positions, providing sufficient temporal grounding without unbounded growth.

\subsection{Data-level transformations} \label{ssec:data_transformations}

Typically ASR is trained using segmented data, allowing models to (unintentionally) exploit boundary effects that may become absent in long-form scenarios (see Section~\ref{ssec:laf}). To encourage the model to leverage explicit signals for acoustic ordering using PE codes (Section~\ref{ssec:pe}) rather than relying on these artifacts, we propose a long-form training approach exposing AD to long-form segments (Definition~\ref{def:lfs}). This differs from other data-level long-form training regimes~\cite{chen2024lf, carvalho25_interspeech}, as we teach AD to specifically decode \lae segments $\calH^s$ in a long stream of audio encodings $\calH$.

\textbf{Expanding acoustic context (AC):} During training with a pair $\{\calY^s, \calX^s\}$, we expand the acoustic encodings to $\calX^s_E = [\calX^s_L,\calX^s,\calX^s_R]$ for randomly selected LF examples to include necessary left/right acoustic contexts for a valid LF segment $\calH^s$ (Definition~\ref{def:lfs}). Lack of edge encodings encourages AD to identify alternative information sources for understanding acoustic encoding order when computing cross-attention, moving beyond reliance on segment boundaries. For AED loss computation and cross-attention, we only use the valid subset of expanded $\calH^s_E$ relevant for $\calY^s$, discarding left $\calH^s_L$ and right $\calH^s_R$ encodings computed for $\calX^s_L$ and $\calX^s_R$, respectively (see shaded regions of Fig.~\ref{fig:lf}).

\textbf{Segment concatenation (SC):} 
LF acoustic recordings allow us concatenate arbitrary numbers of consecutive segments, including their corresponding contiguous acoustic features $\calX$ (with all non-speech audio in the neighbourhood of concatenated segments). We found this to improve AD by i) exposing attention to more diverse segment durations and ii) exposing AD to more \lae encodings. A similar approach has been also recently used in~\cite{carvalho25_interspeech}. SC differs from AC as the latter removes boundary conditions while the former varies training segment lengths, and both can be applied jointly.

\textbf{Semantic segmentation (SS):} In the spirit of using CTC to aid attention decoding ~\cite{moritz2019triggered, HoriMHR21}, and to tie segmentations with decoding~\cite{Huang2022segmentation} we explicitly use CTC head to model segmentation tokens \texttt{\_segE}. These are used to indicate boundaries of semantically coherent sentences, as opposed to relying on audio-only VAD. This tag, when emitted by CTC, is used to determine segments $\calH^s$ for which to run second pass AD. Note that the ability of AD to decode arbitrary subsets of \lae $\calH^s \in \calH$ is one of the main contributions of this work, without which one would need to recompute the encodings $\calH^s$ to introduce boundary cues (see Sections~\ref{ssec:pe} and~\ref{ssec:baselines} to see why). We append segmentation tags for training transcripts using the \textit{Segment any Text} model~\cite{frohmann-etal-2024-segment}.

\section{Experiments}
\label{sec:experiments}

All models in this work are based on the CTC-AED architecture~\cite{watanabe2017hybrid}. It offers appealing deployment flexibility, from streaming to offline or two pass systems. We train two model sizes, referred to as \mbase and \msmall with approximately 90M and 240M parameters, respectively. Both models share the same $R=6$ times convolutional downsampling frontend and attention decoder $\calD_{\theta}$ architecture composed of 3 unidirectional transformer blocks totalling 18M parameters. Acoustic encoders $\calE_{\theta}$ are based on causal Conformer blocks~\cite{gulati2020conformer}\footnote{We use Layer Norm instead of Batch Norm, and RoPE encodings.} 
and have 70M and 219M paremters, respectively. Both share 512 dimensionality and 8 attention heads, with different numbers of blocks and feed-forward (FF) dimensions. \mbase has 12 blocks with 2048 FF units, while \msmall has 28 blocks and 3072 FF dimension. 
%
While architectures like Whisper~\cite{radford2022whisper} distribute parameters equally between encoder and decoder, we scaled the encoder and kept the decoder fixed at 18M params. This allows an efficient decoder for on-device auto-regressive decoding, and a more powerful encoder that better leverages neural accelerators. AD and CTC heads model a distribution over the same 6081 byte-pair subword units~\cite{sennrich2016neural}.

We train our models using a mixture of publicly available data using Adam optimizer for 300k updates, each update using 6144 sentences with an average length of 33 seconds. 
The training data is composed of English corpora accessible through LDC\footnote{https://catalog.ldc.upenn.edu/. We used Fisher, Callhome and ISCI}, 
SpeechOcean, LibriHeavy~\cite{kang2023libriheavy} but also a large-scale LF conversational audio dataset collected from publicly accessible sources, referred to as SpeechCrawl~\cite{gu2025omnirouter}. SpeechCrawl uses pseudo-labels generated using an ensemble of open-source~\cite{radford2022whisper,rekesh2023fast} and in-house models, followed by ROVER committee voting~\cite{Fiscuss97rover}. Not all datasets have long form contiguous acoustic recordings, as such long-form transformations (Section~\ref{ssec:data_transformations}) are applied on LibriHeavy, SpeechCrawl and SpeechOcean. We report our results on public benchmarks including \textit{Tedlium3}~\cite{Hernandez_2018} and \textit{Earnings21}~\cite{delrio2021_earnings} for LF and \textit{CommonVoice}~\cite{ardila-etal-2020-common}, \textit{Librispeech}~\cite{Panayotov2015lbs} for SF decodes, respectively.

\subsection{Results} \label{ssec:baselines}

\begin{table}[!htbp]
\centering
\small
\caption{Impact of proposed modifications on attention decoding on short-- and long-form acoustic encodings (AE). WER (\%) reported on TED-LIUM3 long-form test for \mbase model.}
\label{tab:ablation}
\begin{tabular}{@{}clcccccccc@{}}
\toprule
\multirow{2}{*}{Mdl} & \multirow{2}{*}{AE} & \multicolumn{4}{c}{Modifications} & \multicolumn{3}{c}{WER (\%)} \\
\cmidrule(lr){3-6} \cmidrule(lr){7-9}
 & & SC & AC & PE & SS & \textit{AR} & \textit{AD} & \textit{CAT} \\
 \midrule
 \multirow{2}{*}{0} & \sae & \multirow{2}{*}{\xmark} & \multirow{2}{*}{\xmark} & \multirow{2}{*}{\xmark} & \multirow{2}{*}{\xmark} & 5.8 & 5.1 & 4.9 \\ 
 & \lae & & & & & 5.4 & 295 & 14.5 \\ 
\midrule  \midrule
\multirow{2}{*}{1} & \sae & \multirow{2}{*}{\cmark} & \multirow{2}{*}{\xmark} & \multirow{2}{*}{\xmark} & \multirow{2}{*}{\xmark} & 5.5 & 4.9 & 4.8 \\
 & \lae & & & & & 5.1 & 295 & 16.6 \\
\midrule 
\multirow{2}{*}{2} & \sae & \multirow{2}{*}{\cmark} & \multirow{2}{*}{\cmark} & \multirow{2}{*}{\xmark} & \multirow{2}{*}{\xmark} & 5.6 & 4.9 & 4.8 \\ 
 & \lae & & & & & 5.1 & 40.3 & 8.4 \\
\midrule
\multirow{2}{*}{3} & \sae & \multirow{2}{*}{\cmark} & \multirow{2}{*}{\xmark} & \multirow{2}{*}{\cmark} & \multirow{2}{*}{\xmark} & 5.6 & 5.0 & 4.8 \\
 & \lae & & & & & 5.1 & 145 & 5.5 \\ 
\midrule
\multirow{2}{*}{4} & \sae & \multirow{2}{*}{\cmark} & \multirow{2}{*}{\cmark} & \multirow{2}{*}{\cmark} & \multirow{2}{*}{\xmark} & 5.5 & 4.9 & 4.7 \\
 & \lae & & & & & 5.1 & 5.0 & 4.5 \\  
\midrule
 5 & \lae & \cmark & \cmark & \cmark & \cmark & 4.7 & 4.8 & 4.3 \\
\bottomrule
\end{tabular}
\begin{tablenotes}
\small
\item SC - Segment Concatenation, AC - Acoustic Context, \\ PE - Positional Encoding, SS - Semantic Segmentations \\ AR - Attention Rescoring, CAT - Joint CTC-Attention  
\end{tablenotes}
\vspace{-0.5cm}
\end{table}

\begin{table*}[htbp]
\centering
\small
\caption{WER results for the best systems. Results both on segmented and long-form public benchmarks.}
\vspace{-0.2cm}
\label{tab:final_results}
\begin{tabular}{l|c|c|cccc|cc}
\toprule
\multirow{2}{*}{Model} & \multirow{2}{*}{Decoder} & \multirow{2}{*}{Chunk Size} & \multicolumn{4}{c|}{Segmented} & \multicolumn{2}{c}{Long-form} \\
\cmidrule(lr){4-7} \cmidrule(l){8-9}
& & (seconds) & \textit{Lbs\_clean} & \textit{Lbs\_other} & \textit{CommonV.} & \textit{Tedlium3} & \textit{Tedlium3} & \textit{Earnings21} \\
\midrule
Whisper base.en~\cite{radford2022whisper} & AD & 30 & 4.1 & 9.6 & 17.5 & 4.6 & 4.6 & 12.5\\
\mbase & AR & 0.96 & 2.3 & 5.9 & 16.3 & 4.8 & 4.7 & 13.4 \\ 
\mbase & AD & 0.96 & 2.2 & 5.5 & 14.1 & 4.8 & 4.8 & 13.3 \\ 
\mbase & CAT & 0.96 & 2.1 & 5.4 & 14.4 & 4.4 & 4.3 & 12.2 \\ 
\mbase & CAT & 3.84 & 1.9 & 4.9 & 13.7 & 4.3 & 4.3 & 12.1 \\ \midrule 
Whisper small.en~\cite{radford2022whisper} & AD & 30 & 3.2 & 6.7 & 12.6 & 4.3 & 4.6 & 10.8 \\
\msmall & AR & 0.96 & 2.0 & 4.8 & 13.4 & 4.6 & 4.5 & 12.5  \\ 
\msmall & AD & 0.96 & 2.0 & 4.6 & 12.1 & 4.4 & 4.4 & 12.1 \\  
\msmall & CAT & 0.96 & 1.8 & 4.4 & 12.4 & 4.1 & 4.0 & 11.4 \\  
\msmall & CAT & 3.84 & 1.7 & 3.9 & 11.4 & 3.9 & 3.9 & 11.4 \\  
\bottomrule
\end{tabular}
\begin{tablenotes}
\small
\item AD - Attention Decoding, AR - Attention Rescoring, CAT - Joint CTC-Attention
\end{tablenotes}
\vspace{-0.5cm}
\end{table*}

Table~\ref{tab:ablation} presents a systematic study demonstrating the impact of each proposed modification for fixing AD issues on \lae introduced in Section~\ref{sec:method}. We report results for the three decoding modes - attention rescoring~\cite{yao2021wenet} (\textit{Resc}), Attention~\cite{chorowski2015attention} (\textit{Att}) and joint ctc-attention~\cite{hori-etal-2017-joint} (\textit{CTC-Att}). All three utilize AD in different ways, demonstrating varying behaviour when decoding LF data. Unless stated otherwise, segmentations for \sae and \lae are identical and derived from a stand-alone VAD system for Models 0 - 4. The results show a stark contrast between performance on \sae and \lae, highlighting the challenges of applying AD to transcribe LF audio segments.

\textbf{Baseline Performance (Model 0)} 
The first block of Table~\ref{tab:ablation} reports word error rates (WER) for the baseline AED model. We can see that i) there is a significant accuracy gain when going from attention rescoring to autoregressive decoding (5.8\% to 4.9\%) for \sae, and ii) going from \sae to \lae is also beneficial for attention rescoring (5.8\% to 5.4\%, due to better quality of first pass CTC hypotheses). However, there is a significant degradation for \lae with attention or CTC attention decoders. This demonstrates that vanilla AD trained on segmented data is incompatible with \lae representations (Definition~\ref{def:lfa}). Closer investigation showed that the model struggled to emit EOS token, which resulted in re-decoding same chunks over and over again causing excessive insertions (only limited by decoder setting of max number of tokens). 

\textbf{Segment Concatenations (SC, Model 1)}
Segment concatenation is the first modification we introduce to increase AD performance on LF data, by training it on longer multi-segment utterances (up to 2 minute 30 seconds in this work). We can see this approach decreases general WER across the board for all decoding modes for \sae, though does not solve the crucial AD failure for \lae segments, \ie the segments with no boundary cues preserved (Definition~\ref{def:lfs}). 

\textbf{Impact of Acoustic Context (AC, Model 2)}
Introducing acoustic context during training maintains segmented performance while  mitigating the baseline's difficulties on \lae, reducing WER from 295\% to 40.3\%, still a large performance gap when compared to \sae. The extended AC encourages the model to rely less on boundary cues and more on the existing relative positional information present in acoustic embeddings via the existing input RoPE~\cite{su2024rope} positional encodings. For \mbase model, $L=12, N=16, M=16$ and $R=6$, so $C_L^{\max}=1152$ and $C_R^{\max}=96$ acoustic frames. To obtain \lae, we append thus 11.52s and 0.96s seconds of left and right context, respectively (see also Section~\ref{ssec:laf}).

\textbf{Positional Encoding Effects (PE, Model 3)}
Adding auxiliary absolute positional encodings to $\calH^s$ (Model 3) shows modest improvements, roughly halving long-form AD WER to 145\% compared to Model 1. 
Although less effective than acoustic context, positional encoding should provide essential temporal structure information that helps AD make sense of sequence ordering in long-form audio. Interestingly, appending these is not sufficient for the model to fully leverage them, likely due to boundary cues in segmented training data being an easier shortcut to learning temporal ordering. Note, however, that these auxiliary position codes do improve over existing Rope encodings already present in the inputs (\cf Model 1).

While the results for attention-only decoding for Model 3 suggest that AC expansion (Model 2) produces larger WER reductions than segmental PE codes (Model 3), a deeper analysis reveals more nuanced differences between these approaches. Positional codes (Model 3) enable the attention mechanism to decode more coherent and longer sentences, but the model struggled to emit EOS tokens. In contrast, acoustic expansion (Model 2) generated EOS tokens more frequently, though at the cost of increased deletion errors.

These findings suggest that Model 3 could achieve superior overall performance if provided with better guidance on when to stop decoding. We tested this hypothesis using joint \textit{CTC-Attention} decoding~\cite{hori-etal-2017-joint}, with results shown in the last column of Table~\ref{tab:ablation}. When guided by CTC, attention decoding with auxiliary positional codes (Model 3) demonstrates better accuracy than the model with expanded acoustic context during training (Model 2), 5.5\% vs 8.4\% WER, respectively.

\textbf{Combined Context and Positional Encoding (Model 4)}
The combination of acoustic context and positional encoding (Model 4), though, yields significant improvements, achieving parity on long-form audio w/ AD between \sae and \lae encodings. This result suggests that both spatial (acoustic context) and temporal (positional encoding) information are complementary and necessary for long-form attention decoding, and that exposing the model to more data without boundary cues during training encourages it to rely more on auxiliary positional encodings. This result also demonstrates that explicit PE information added to $\calH^s$ is important (compare Table~\ref{tab:ablation} \textit{Att} decode results of Models 4 and 2).

\textbf{Complete System with all modifications (Model 5))} Model 5 demonstrates that using first-pass CTC semantic segmentation tokens to trigger attention decoding consistently outperforms VAD-based approaches. This improvement is expected since the attention decoder is trained primarily on correctly segmented utterances, while VAD may introduce non-semantic boundaries that create inference mismatches. In our long-form experiments, these VAD inconsistencies typically resulted in increased deletion errors.

The CTC-attention hybrid decoder achieves the best overall performance across all decoding modes. Notably, Model 5 operates naturally in the \lae space, with Model 4 serving as its corresponding \sae baseline when using VAD segmentation. Similar performance trends were observed on the Earnings-21 long-form test set, though these results are not reported here.

\subsection{Final results}

Table~\ref{tab:final_results} demonstrates that our proposed system achieves strong performance across both segmented and long-form evaluation datasets. Importantly, the adaptations made to enable effective long-form processing do not degrade performance on standard segmented audio, indicating our architectural modifications are robust to different types of segmentations.

Parameter-wise our models deliver comparable performance to similarly-sized Whisper variants, while doing so at lower latency. Our model matches Whisper base.en performance on long-form datasets while improving on segmented tasks, and our small model consistently outperforms Whisper small.en across most benchmarks. Note that \textit{Librispeech} \textit{lbs\_clean}, \textit{lbs\_other} and \textit{CommonVoice} are zero shot for Whisper, but not for our systems, which could explain better performance of our models across these three benchmarks.

The use of CAT decoding provides additional practical advantages. This hybrid approach often yields the best performance (e.g., 1.8\% WER on \textit{lbs\_clean} vs. 2.0\% with pure attention), while also enabling streaming first pass inference capabilities, semantic segmentation tags for when to trigger AD, and providing more robust decoding in challenging acoustic conditions where attention-based decoding is prone to hallucinations. Since our model was trained with variable masking at the encoder level, we can configure the model's chunk size to a larger value. We report these results in Table~\ref{tab:final_results} for chunk size of 3.96s for the best CAT decodes. The overall trend is WER gets further reduced, as expected, but at higher latency cost.

\section{Conclusions}
\label{sec:majhead}

We presented a set of modifications to train attention decoder models so that they can process long-form acoustic encodings. Through acoustic context expansion, auxiliary positional encoding, and long-form segment sampling and concatenation,  the attention decoder based system achieves transcription quality of long-form acoustic encodings that is on par with decoding of segmented encodings, enabling new ways to use attention decoders. The hybrid CTC-attention system with first pass semantic segmentation tags provides a low latency approach to use AD two-pass architecture, when compared to models that need to operate on fixed size input windows or raw signal based voice-activity-detection decisions. While we enabled AD to work seamlessly with long-form acoustic encodings, we did not investigate additional ways to extend AED with contextual continuity between different segments~\cite{hori20_interspeech,radford2022whisper}. We leave this effort for the future work. Described advancements helped to improve a number of user-facing features such as Live Translation, Call Screening, Hold Assist, Live Voicemail, Visual Voicemail \& Call Transcription, FaceTime Live Captions \& System-wide Live Captions, Audio Transcription in Notes, Voice Memos \& Journal and Transcribe to Captions in Final Cut Pro.

\section{Acknowledgements}

We thank Nikos Flemotomos, Erik McDermott and Barry Theobald for useful suggestions on this work. Thanks to Shaoen Qin for large scale SpeechCrawl pseudo-labelling.

\bibliographystyle{IEEETran}
\bibliography{bib/e2e,bib/aed}

\end{document}